\documentstyle[referee,epsfig]{mn}
\newif\ifAMStwofonts
\newcommand{\be}{\begin{equation}}
\newcommand{\ee}{\end{equation}}
\newcommand{\bea}{\begin{eqnarray}}
\newcommand{\eea}{\end{eqnarray}}

\newcommand{\etal}{ {\it et al.} }

\title{Coarse Graining the Distribution Function of Cold Dark Matter  }

\author[ R.N Henriksen \& M. Le Delliou]
{
Richard N. Henriksen$^1$\thanks{henriksn@astro.queensu.ca}, and
M. Le Delliou$^1$\thanks{delliou@astro,queensu.ca}\\
$^1$Queen's University, Kingston, Ontario, K7L 3N6,Canada \\
}
\date{\today}
\pagerange{\pageref{firstpage}--\pageref{lastpage}}
\pubyear{2001}

\begin{document}

\maketitle

\label{firstpage}

\begin{abstract}
Many workers have found that the recollapse of a dark matter halo
after decoupling has a self-similar dynamical phase. This behaviour is
maintained strictly so long as the infall continues but it appears to evolve
smoothly into the virialized steady state and to transmit some of its
properties intact.
 The density profiles established in this phase
are all close to the isothermal inverse square law however, which is steeper than the
predictions of some n-body
simulations for the central regions of the halo, which are in turn steeper
than the density profiles observed in the central regions of some
galaxies; particularly dwarfs and low surface brightness galaxies. The
outer regions of galaxies both as observed and as simulated have
density profiles steeper than the self-similar profile. Nevertheless there appears to be an intermediate region in
most galaxies in which the inverse square behaviour is a good
description. The outer deviations can be explained plausibly in terms
of the transition from a self-gravitating extended halo to a Keplerian
flow onto a dominant central mass (the isothermal distribution can not
be complete), but the inner deviations are more
problematic. Rather than attack this question directly, we use in this
paper a novel coarse-graining technique combined with a shell code to
establish  both the distribution function associated with the
self-similar density profile and the nature of the possible deviations
in the central regions. In spherical symmetry we find that both in the
case of purely radial orbits and in the case of orbits with non-zero angular
momentum the self-similar density profile should flatten progressively
near the centre of the system. The NFW limit of $-1$ seems possible.
In a section aimed at demonstrating our technique for a spherically
symmetric 
steady state, we argue that a Gaussian distribution function is the
best approximation near the centre of the system.      
     
\end{abstract} 
\setlength{\baselineskip}{13pt}
\section{Introduction}
\label{sec:intro}
The study of the radial infall of dark matter and the realization that
the evolution becomes self-similar has by now a long history (see for
example \cite{HW99} for a summary of the relevant
references and a nearly state of the art statement of our understanding ). The salient theoretical behaviour is well known if not
completely understood, but it yields density profiles that are
theoretically distinct from the inverse square law but observationally
indistinguishable from this law. However the NFW law from n-body
simulations \cite{NFW} even if slightly steeper
when resolution effects are taken into account (e.g. \cite{Moore98}; but see also \cite{Kravtsov98}) is distinctly flatter than the inverse square law in the central
regions. Moreover in
some galaxies (low surface brightness or LSB galaxies, DE Blok \etal
2001; some dwarf ellipticals, Stil 1999; and even brighter elliptical
galaxies compared to the fainter objects, Merritt and Cruz 2001) the
observed density profile is significantly flatter in these regions
than even the n-body profiles (except \cite{Kravtsov98}). Both
observed and simulated profiles are steeper than the inverse square
law in the outer regions, but this can be understood \cite{HW99} in terms of secondary accretion after most of the halo
mass has fallen in. 

We take the view in this paper that if finite resolution effects are
at work in the centre of the simulations, and perhaps also in reality since obviously
real particles can not yield an infinite density cusp; then we may
gain some insight by solving analytically the Collisionless Boltzmann Equation
(CBE) for the distribution function (DF) in an series expansion in powers of the inverse `smoothing length' (to be defined
precisely below). In such an expansion the lowest order is the
coarsest grained approximation and higher orders yield progressively
finer grained information. It transpires that in this way we can
deduce the radial distribution function necessary to maintain the
strict self-similarity and we may place some constraints on
modifications to this DF that can produce flattening.

Our method of coarse graining is by non-canonical transformation on
the phase space (we make a special choice of a stretching
transformation here, but the method is more
general) which produces a definite equation to be satisfied by each
term in our series. We illustrate the method first by applying a
stretching transformation to a spherically symmetric isotropic
equilibrium. The bulk of the paper is devoted to exploring the
self-similar `equilibrium'maintained by the continuing accretion onto a relaxed
core \cite{FG84}; \cite{B85}; \cite{HW99}.

 A consideration of the entropy for the system
allows us to suggest that the central DF in spherical equilibrium is
Gaussian in velocity with a separable radial dependence, while in
exact self-similar secondary infall the DF is everywhere proportional
to the square root of the particle binding energy. 
Moreover the requirement that the coarse grained
entropy be maximal predicts perturbations that flatten the inner
regions of steep self-similar profiles and steepen the outer regions
of flat  self-similar profiles. We show also that the addition of
angular momentum to spherically symmetric orbits does not change these conclusions.
       
\renewcommand{\textfraction}{0}
\renewcommand{\topfraction}{1} 
\renewcommand{\bottomfraction}{1}

\section{Coarse Graining of the CBE by Non-Canonical Phase Space Transformation}
\label{sec:Coarse Grain}

We wish to coarse-grain the CBE together with the mean field Poisson
equation  in some systematic fashion. 
A simple smoothing of phase space to get a coarser grid over which to
define the DF suffers from non-commutation with the operators in the
CBE and leads to a coarse grained DF that is defined only
qualitatively (e.g. \cite{BT87}). Our idea is to use a
coordinate transformation on phase space that does not preserve phase
space volume and so is non-canonical. Although this can be applied to
evolving systems it is not very well defined because of a lack of
information concerning the initial conditions of the coarse grained
(time averaged for an evolving system) DF. Thus we will restrict
ourselves to systems that have attained `equilibrium' in some
appropriate variables (see below). 

The simplest example of this
idea is to apply a stretching transformation to phase space, and in
fact this is the only transformation that we have explored (the scale
factor depends on time in the self-similar examples). However more
general non-canonical transformations may be found that are useful so
we continue with this label. 

This can
be readily carried out on an equilibrium spherical collisionless
system  for which the governing equations are;
\be 
v\partial_r f-\partial_r\Phi\partial_v f=0,
\label{eq:CBEiso}
\ee
 and 
\be
\partial_r(r^2\partial_r\Phi)=4\pi G\rho~r^2\equiv 4\pi G r^2 \int~f~4\pi
 v^2 dv,
\label{eq:Poissoniso}
\ee

where $f$ is the normal phase space mass density and $v$ is the
magnitude of the isotropic velocity. A scaling or
stretching transformation can be applied to the phase space as 
\bea
r\leftarrow r\ell&   & v\leftarrow uv,\\
f\leftarrow f\mu &   & \Phi\leftarrow 4\pi G \mu\ell^2 u^3
\Phi,\\
\rho\leftarrow \mu u^3\rho,&   & \label{eq:scale}\eea
where $\ell$, $\mu$ and $u$ represent constant factors that we may 
take here to be dimensionless numbers. The phase space volume
thereby transforms by the factor $\ell^3 u^3$ and we will take $\mu$
equal to the reciprocal of this volume times a fiducial mass $m$ in order to conserve mass (for
unbounded systems). Thus $\ell$ and $u$ are the theoretical smoothing
`lengths' on phase space. Finally we scale $f$ according to
$f\leftarrow f/(4\pi G)$.

The equations are the same in the stretched variables except that the
CBE becomes
\be
v\partial_r f-{\cal R}^{-1}\partial_r\Phi\partial_v f=0,
\label{eq:stretchCBE}
\ee
where 
\be 
{\cal R}\equiv u^2\ell/4\pi Gm.
\ee
As either or both of the phase space variables are stretched this
parameter becomes large.

We  coarse grain the system by assuming a convergent
expansion of the form 
\be 
f=\Sigma _{i=0} f_i(r,v){\cal R}^{-i},
\label{eq:series}
\ee
so that the lowest order contains the least information about the
DF. A similar expansion follows for the density and the potential. 
     
Proceeding in this fashion we find first that $f_o=f_o(v^2)$ and
consequently that
\be 
\rho_o= \int~f_o 4\pi v^2 dv.
\ee
The limits in this equation would normally be set so as to include
just the  bound particles. However the stretched specific energy
($E\leftarrow u^2 E$) is related to the other stretched quantities as 
$E =v^2/2+ {\cal R}^{-1}\Phi$ and so is purely kinetic to lowest
order. Thus we take the upper limit for $\rho_o$ to be simply $v_m$,
some maximum speed at $r=0$, and the lower limit to be zero. The integral so
defined is denoted $\rho_o(v_m)$. Consequently one finds the harmonic
potential in lowest order (so that all particles are bound and turn
where $E={\cal R}^{-1}\Phi$) 
\be 
\Phi_o=\rho_or^2/6. 
\ee

The first order term in the DF expansion is now found to be 
\be 
f_1=\frac{\rho_o r^2}{3}\frac{df_o}{dv^2},
\ee
and consequently 
\bea
\rho_1&=& \frac{4\pi}{3}\rho_or^2~\int_0^{v_m}~v^2\frac{df_o}{dv^2}dv,\\
\Phi_1&=& \frac{\pi}{15}\rho_or^4~\int_0^{v_m}~v^2\frac{df_o}{dv^2}dv,
\label{eq:first order}
\eea
assuming no central point mass. The next order terms in these series may
be found to be proportional to $r^4$, since for example
\be
\partial_rf_2=\frac{2\rho_o^2
  r^3}{9}\frac{d^2f_o}{d^2v^2}+\frac{2}{5}\rho_1~r\frac{df_o}{dv^2},
\label{eq:2nd order}
\ee
 and so on. 
In this way we are
generating an expansion of the type necessary in general for
polytropes (e.g. \cite{Chandra57} p.94), but here we are able to
relate the coefficients directly to the DF that is not necessarily a polytrope. In essence since most of the physical
system will be at small r for an arbitrarily stretched coordinate, we
are able to re-interpret the polytropic expansions in terms of
 progressively finer graining.  

The second order term in the DF is proportional to $r^4$, but more
importantly it involves the second derivative of $f_o$ with respect to
$v^2$.
This behaviour will continue to all orders where the highest order 
derivative of $f_o$ appearing is equal to the coarse-grained
order. One must therefore be careful to ensure that these derivatives
are sufficiently well-behaved to all orders that a divergent density
is avoided, which would render the expansion invalid. For example if
we try to imitate a polytrope by setting (A is a positive constant)
$f_o=A(\psi+v_m^2-v^2)^{(n-3/2)}$, then all derivatives are
well-behaved for $v\le v_m$ at the cost of a discontinuity in the DF
at $v_m$. The quantity $\psi$ appears as an negative constant
potential (which is stretched by the factor $u^2$) to be added to $\Phi_o$ and in principle we can keep only
the negative energy particles by setting $v_m(r)=\sqrt{2(\psi-{\cal
    R}^{-1}\rho_o r^2/6)}$. However the presence of ${\cal R}$ in this
expression breaks the ordering of our
expansion and although it may improve the rapidity of convergence, we
continue to treat $v_m$ as a constant and leave the corrections to
higher orders. 

To obtain a continuous transition to zero at $v_m$, we might take for
$f_o$ 
one of:
a Gaussian DF   
\be
f_o=A e^{(v_m^2-v^2)/2\sigma}~~;
\label{eq:isothermal}
\ee
 or a King type DF (a constant potential is absorbed into $v_m^2$)
\be
f_o=A(e^{(v_m^2-v^2)/2\sigma}-1).
\label{eq:King}
\ee

Ultimately the choice of $f_o$ must be that which best reflects the
coarse-grained image of the system in question. This seems to require us to
know something about the gross properties of the system `a priori` so that in fact
our procedure is more like a progressive `fine graining'. Each choice
of coarse-grained DF $f_o$ will generate by this procedure (provided
that it may be carried out properly as discussed above), a
progressively more precise DF as the fine graining terms are added.
 
We may seek general principles concerning the form of $f_o$ however. We note that the choice of the
Gaussian is unique in the present context in that it will maintain this dependence in the DF
to all orders. In addition it will maximize a version of the statistical entropy \cite{N2000},\cite{LB67}. The expansion is well defined to all orders for the Gaussian, but we can only expect it to apply to the central regions of
finite systems (otherwise one will not conserve mass as the scale is
stretched).

The expansion of the Gibbs' statistical entropy (essentially the
generalized `H function' of Boltzmann when there are no correlations) in the form ($f_o$ is
normalized by the total mass of the system) 
\be
S=\int~ d\tau{f_o\ln{1/f_o}-\frac{f_1}{{\cal R}}(1-\ln{1/f_o})+\cdots}, 
 \label{eq:entropy}
\ee
shows that since $f_o$ is a small number (non-degenerate system) and 
since $f_1<0$ ( actually true for all three cases mentioned above
although we require $n>3/2$ for the polytrope), the
entropy decreases as the fine graining terms are added. This seems as
it should be. Moreover with the choice of the  Gaussian for $f_o$ the
DF is separable. This makes the velocity distribution the same at all
$r$ as one might expect in the central equilibrium region. 

Our conclusion in this section is that the best behaved coarse
  graining expansion is consistent with the proposition that  
  {\it a Gaussian should be the DF in the central regions of
  finite spherical systems that have undergone relaxation and hence
the evolution towards maximum entropy}. Our arguments do not of course
  prove this proposition, but they support the statistical arguments
  \cite{N2000},\cite{Shu78},\cite{Shu87}, \cite{LB67}. We have
  essentially used the idea that in a relaxed system we expect the
  fine grained DF and the coarse grained DF to be the same. That is
  that the DF should be independent of cell size. In fact the principal
  advantage of our method is that the coarse graining expansion
  dispenses with the necessity in the combinatorial statistical treatment
 of making an arbitrary distinction between microcells and macrocells
  \cite{Shu78}. Moreover it is in agreement with the maximum entropy
  approach of Nakamura \cite{N2000}. 

It may seem contradictory to infer a DF
  that remains the same at all radii as found above while also restricting
  our proposition to the central regions of the system. In fact in a
  completely relaxed system the DF should be the same at all radii
  as in fact is true for the Gaussian DF. But we know this to yield an infinite
  system \cite{LB67}, and so presumably deviations from the relaxed
  state increase with radius in any finite system. This means that at large radii the
  coarse grained DF and the fine grained DF no longer coincide and the
  coarse Grained DF need no longer satisfy Liouville's theorem. Moreover our stretching transformation
  can not be applied too close to the boundary of a finite system.  

These ideas suggest for example that polytropes
are not in a maximum entropy condition but rather have to be
constructed rather carefully. The density profile consistent with a
Gaussian DF 
 does not have a central cusp so that simulations that
predict such cusps are presumably not sufficiently relaxed near the
centre. Our investigations of the next two sections suggest in fact
that progressively finer resolution of the central regions of
self-similar infall produce flatter density profiles.

\section{Coarse Graining of Self-Similar Infall with Radial Orbits}
\label{sec:SScoarse grain}
When we restrict ourselves mainly to radial orbits, we
will introduce as usual the canonical distribution function
$F(t,r,v_r)$ such that the complete phase space mass density is given
by 
\be 
f\equiv (4\pi^2G)^{-1}\delta (j^2)~F,
\ee
where $j^2\equiv r^2(v_\theta^2+v_\phi^2)$.
The mean field treatment is then given by the combined CBE and Poisson
equation in the forms(see e.g. Henriksen and Widrow, ibid)
\be 
\partial_t F +v_r\partial_r F -\partial_r\Phi\partial_{v_r}F=0,
\label{eq:CBErad}
\ee
and 
\be
\partial_r(r^2\partial_r\Phi)=\int~F~dv_r=4\pi G r^2\rho,
\label{eq:Poissonrad}
\ee
where $\Phi(t,r)$ is the gravitational potential and $\rho(t,r)$ is
the mass density.

In the previous section we pointed out that our procedure made the
most sense when applied to a system that is in `equilibrium' since
otherwise the coarse graining would also be over the history of the
system. However a time-dependent case which is nevertheless in a kind
of equilibrium is the self-similar phase of the formation of the
system. In fact by changing to variables appropriately scaled by the
time \cite{FG84}; \cite{B85}; \cite{HW97};\cite{HW99} the equation is transformed to an equivalent steady
problem. 

Here we adopt these variables in a formalism given by Carter
and Henriksen (1991; see also \cite{HW95}; \cite{HW99}; \cite{H97}) where
we adopt an exponential time $T$ 
\be 
e^{\alpha T}=\alpha t,~dT/dt =e^{-\alpha T}.
\label {eq:time}
\ee
It is convenient to take time to be measured in units of some
fiducial value so that $\alpha$,the stretching scale for time, is also
a dimensionless number. We introduce the scaled phase space variables
$X$, and $Y$ for the radius and radial velocity respectively as (in
the previous section the scaled quantities were not given new labels
but we continue to do this here to agree with previous practice) 
\be
 X\equiv re^{-(\delta/\alpha)\alpha T},~~Y\equiv v_re^{-(\delta/\alpha-1)\alpha T},
\label{eq:phasesp}
\ee
where $\delta$ provides for an independent spatial stretching
scale. However in a self-similar system this ratio is constant. In addition we scale
the radial canonical phase space DF according to  
\be
P(X,Y)\equiv Fe^{-(\delta/\alpha-1)\alpha T},
\label{eq: radDF}
\ee
where the lack of a dependence on $T$ in the scaled DF $P$ is the
condition for self-similarity. The potential is correspondingly scaled
to $\Psi(X)$ where 
\be 
\Psi\equiv e^{-2(\delta/\alpha-1)\alpha T}\Phi,
\label{eq:radpot}
\ee
and the density is scaled to $\theta(X)$ where 
\be
\theta\equiv \rho e^{2\alpha T}.
\label{eq:radrho}
\ee
Consequently 
\be 
\theta=\frac{1}{4\pi X^2}\int~P~dY.
\label{eq:scaledens}
\ee
We observe that the phase space volume element $\Delta r\Delta v_r$ has been scaled
according to 
\be 
\Delta r\Delta v_r=\Delta X \Delta Y e^{(2\delta/\alpha-1)\alpha T}.
\label{eq:volelement}
\ee
 The scaling `preserves' mass since 
\be
P\Delta X \Delta Y e^{(3\delta/ -2\alpha)T}\equiv F\Delta r\Delta
 v_r\equiv \Delta m ,
\ee
together with $3\delta-2\alpha=\mu$ where $\mu$ is the mass scale
 (e.g. \cite{HW95}) give
 the correct time dependence for the mass element $\Delta m$ during the
 self-similar phase (which is only constant for Keplerian
 self-similarity wherein $\delta/\alpha=2/3$).

Equations (\ref{eq:CBErad}) and (\ref{eq:Poissonrad}) now become
respectively 
\bea
(\frac{\delta}{\alpha}-1)P+\left(\frac{Y}{\alpha}-(\frac{\delta}{\alpha})X\right)\partial_XP&-&\left((\frac{\delta}{\alpha}-1)Y+\frac{1}{\alpha}\frac{d\Psi}{dX}\right)\partial_YP=0,\\
\frac{d}{dX}\left(X^2\frac{d\Psi}{dX}\right)&=&\int~P~dY.
\eea
These are the equations to be coarse grained in this case.

We make use of the parameter $\alpha$ in the
transformation to the self-similar phase space variables to effect the
coarse graining. We note that equation (\ref{eq:volelement}) shows
that a fixed volume element $\Delta X\Delta Y$ at a fixed time
corresponds to an ever larger volume element $\Delta v_r\Delta r$ as
$\alpha$ is increased while holding the ratio $\delta/\alpha$ constant
so as to maintain the similarity class. This is true so long as the
similarity class is $>1/2$, which is usually the case of interest
since the Keplerian value of $2/3$ tends to be the minimum value
encountered. The value $\delta/\alpha =1/2$ gives a canonical
transformation by equation (\ref{eq:volelement}) and no change in the
volume of phase space is effected. Increasing $\alpha$ amounts to a parametric way of
stretching time and space so that it becomes the theoretical smoothing
length parameter. Normally we take $\alpha=1$ which can be regarded as the
fine grained limit. Here however we use the freedom in its absolute
value to allow it to adopt large values and thus yield a coarse grained
limit (with the similarity class fixed the spatial scale is also being
stretched in proportion).

The expansion we use is again of the form 
\be
P(X,Y)=\Sigma_{i=0}P_i\alpha^{-i},
\label{eq:expansion}
\ee
and similar expansions apply to the density and the potential. We
begin with the zeroth order equations which become (it is easy to
write the nth order formally but this is not very transparent)
\bea
(\frac{\delta}{\alpha}-1)P_o-(\frac{\delta}{\alpha})X\partial_XP_o
&-&(\frac{\delta}{\alpha}-1)Y\partial_YP_o=0,\\
\frac{d}{dX}\left(X^2\frac{d\Psi_o}{dX}\right)&=&\int~P_o~dY.
\eea

We readily find $P_o$ by the method of characteristics to be 
\be 
P_o=P_{oo}(\zeta)X^{(1-\alpha/\delta)},
\label{eq:zeroP}
\ee
where $P_{oo}$ is an arbitrary function of $\zeta$, which is constant on
characteristic curves of $P_o$. These curves are in turn given by 
\be
\zeta=\frac{Y}{X^{(1-\alpha/\delta)}},
\label{eq:charzero}
\ee
and the actual arc-length $s$ of a characteristic in phase space may be
taken to be 
\be
s=\frac{\alpha}{\delta}\ln{X},
\label{eq:arc}
\ee
and this can be used to give $X(s)$, $Y(s)$ if desired.

If we recall equation (\ref{eq:scaledens}) and combine it with
(\ref{eq:charzero}) for $dY$  and (\ref{eq:zeroP}) for $P_o$  then we may write 
\be
\theta_o= \frac{X^{-2\alpha/\delta}}{4\pi}\int~P_{oo}(\zeta)~d\zeta.
\label{eq:denszero}
\ee

Thus the zeroth order coarse graining already produces the
self-similar density profile in $r$, namely $r^{-2\alpha/\delta}$ (cf
\cite{HW95}; \cite{HW99}). This is not surprising once
the self-similarity is imposed since it is essentially a dimensional
argument (e.g. \cite{HW95}). Since however at this level no
potential enters the calculation, it does suggest that such elements
of a simulation as smoothing length may not be essential to finding
this profile if the other elements that impose the self-similarity
(such as initial density profile) are in place. We shall see in a
shell code application below that there is some evidence for this.

The zeroth order potential is 
\be
\Psi_o=\frac{ I_{oo} X^{(2-2\alpha/\delta)}
  }{(3-\frac{2\alpha}{\delta})(2-\frac{2\alpha}{\delta})}\equiv 
-\gamma X^{(2-2\alpha/\delta)},
\label{eq:potzero}
\ee
where $I_{oo}$ is the integral over $P_{oo}$ occuring in equation
(\ref{eq:denszero}) and $\gamma$ is defined to be positive when
$\alpha/\delta>1$. The cases $\alpha/\delta=1$ and $\alpha/\delta=3/2$
are logarithmic rather than power law and must be treated separately.
The potential is not in general simply harmonic (except in the steady case where
$\alpha=0$ that does not concern us here. An expansion in positive
powers of $\alpha$ would allow a `fine graining' about this state.)

Finally at this order we note that the energy of a particle scales as 
$E=e^{2(\delta-\alpha)T}\epsilon$ where 
\be
\epsilon=Y^2/2+\Psi(X),
\ee
and with $\epsilon_o=\epsilon_{oo}X^{2(1-\alpha/\delta)}$ we find that
$\zeta$ is related to the zeroth order energy through 
\be
\zeta =\pm\sqrt{2(\epsilon_{oo}+\gamma)}. 
\label{eq:zeta}
\ee
Consequently  the limits of the integration over
$\zeta$ at this order are $[-\sqrt{2\gamma},+\sqrt{2\gamma}]$ when
$\gamma$ is positive in order to include only the bound particles,
while when $\gamma<0$ we have a case similar to the harmonic potential
of the previous section and $\zeta$ varies between plus or minus some
maximum value.

We now proceed to the first order terms for which the following
equations must be solved
\bea
(\delta/\alpha-1)P_1-\frac{\delta}{\alpha}X\partial_XP_1-(\delta/\alpha-1)Y\partial_YP_1&=&-Y\partial_XP_o+\frac{d\Psi_o}{dX}\partial_YP_o,\\
\frac{d}{dX}\left(X^2\frac{d\Psi_1}{dX}\right)&=&\int~P_1~dY.
\label{eq:firstorder}
\eea
Once again the method of characteristics yields a solution as 
\be
P_1=X^{(1-2\alpha/\delta)}\left((\alpha/\delta-1)\zeta(P_{oo}-\zeta P'_{oo})+I_{oo}P'_{oo}/(3-2\alpha/\delta)\right)\equiv
P_{11}(\zeta)X^{(1-2\alpha/\delta)},
\label{eq:firstDF}
\ee
where the shape of the characteristics (for $P_1$) is the same as
 for $P_o$ above and the prime indicates differentiation with respect
 to $\zeta$. In addition the form for $\theta_1$ follows as
\be
\theta_1=\frac{X^{(-3\alpha/\delta)}}{4\pi}\int~P_{11}(\zeta)~d\zeta.
\label{eq:thetafirst}
\ee  
Thus the solution to first order in the coarse graining parameter
(i.e. a finer grained solution) takes the form 
\bea
P&=&X^{(1-\alpha/\delta)}\left(P_{oo}(\zeta)+\alpha^{-1}P_{11}(\zeta)X^{-(\alpha/\delta)}\right),\\
\theta&=&\frac{X^{-(2\alpha/\delta)}}{4\pi}\left(
  I_{oo}+\alpha^{-1}X^{-\alpha/\delta}I_{11}\right),
\label{eq:1stcoarse}
\eea
where $I_{11}$ is the integral occurring in $\theta_1$, which is not
however positive definite in this case.

Now what is remarkable about these forms is that they suggest that at
small $X$ (large $t$ and/or small $r$) there are deviations from the
coarse grained self-similar behaviour as we increase the resolution or
information content. Indeed should $P_{11}$ be negative then there
would be a flattening of the density profile in this limit. However we
know from the numerical simulations that there is a region where the
self-similar behaviour is maintained so long as the infall
continues. This leads us to ask a question similar to that which was
posed in the previous section. There we asked for separability in the
DF so as to yield the same velocity distribution at all radii. Here we
can ask for the DF which maintains the self-similar density profile
for all $X$. This requires $P_{11}=0$ since then all higher order tems
in the expansion will vanish and the density profile is uniquely
self-similar. We can write this term from equations (\ref{eq:firstDF})
and (\ref{eq:zeta})
in the form 
\be
P_{11}=(\alpha/\delta-1)\zeta
P_{oo}\left(1-2\frac{d\ln{P_{oo}}}{d\ln{\epsilon_{oo}}}\right),
\label{eq:P11}
\ee

so that the condition that all higher order terms vanish is simply 
\be 
P_{oo}=const\times |\epsilon_{oo}|^{1/2}\equiv const \times
|\zeta^2/2-\gamma|^{1/2}.
\label{eq:SSDF}
\ee

Retracing the various definitions and scalings, we can discover that
this last result gives for the canonical radial DF 

\be
F(E)=const \times |E|^{1/2},
\label{eq:canonDF}
\ee
and hence the full DF is 
\be
f= const.\times\left(4\pi^2\right)^{-1}\delta(j^2)|E|^{1/2}.
\label{eq:radDF}
\ee
This tends to confirm the conjecture of \cite{HW99}.
This was based largely on the argument for a continuous transition to the steady
state (a steady state found also in \cite{HW95}) together with some
weak evidence from a shell code simulation. Our present argument is
related since in the limit that $\alpha\rightarrow\infty$ we obtain a
steady state, and in fact it is readily verified from
(\ref{eq:charzero}) that $\zeta$ is independent of time. By requiring
all higher orders to vanish we have effectively selected this steady state.

In the next section we shall show some increased numerical evidence
for this DF. The evidence is best when a point mass is allowed
to be present at the centre, even if it is a negligible fraction of
the total halo mass. This might be termed a `black hole' but
in reality it simply imitates the density singularity expected from the
self-similar density profile. That profile does not yield a finite
mass at the centre but it does predict a cusp that the numerical work has
difficulty expressing in the absence of a central point mass. When the
point mass is small compared to the halo mass it serves to imitate a cusp
without a finite mass singularity. Thus we
might well expect the DF in this case to be closest to the $(-E)^{1/2}$
behaviour (see below).
   
The real significance of this result is to show that deviations from
the self-similar infall profile of radial orbits in spherical symmetry (a subsequent
section will deal with more general orbits) will be
expressed as deviations from the $(-E)^{1/2}$ law. These may well be due to incompleteness of the
self-similarity at the centre, to angular momentum or to geometric
effects. We have already seen that an abrupt cut-off at high binding
energies does create a deviation in the density profile, but only to
logarithmic order \cite{HW99}. More global changes
must occur in the DF in order to produce substantial changes in the
density profile even near the centre.

 Finally it is also of interest to consider the entropy expansion. We
 follow custom (e.g. \cite{BT87}) in writing the Gibbs'
 entropy as 
\be 
S=-\int~f\ln{f}~d\tau,
\ee
where $d\tau=4\pi r^2~dr~ 2\pi j dj~dv_r/r^2$. 
This is the generalized (i.e. to inhomogeneous systems)
 H function of Boltzmann in the absence of particle-particle
 correlations. Then if we scale the
 entropy  and mass of the system according to 
\bea
{\cal S}&=&S~e^{-(3\delta-2\alpha)T},\\
{\cal M}&=&M~e^{-(3\delta-2\alpha)T},
\eea
we may find that 
\be 
{\cal S}={\cal M}(\ln(4\pi^2)+(1-\delta/\alpha)~\alpha T)-\int~dX~dY~P\ln{P},
\label{eq: entropy}
\ee
where the scaled mass is related to the scaled density in the natural
way 
\be 
{\cal M}\equiv \int~4\pi X^2~\theta~dX.
\ee
Thus the only time dependence in this self-similar entropy is the
explicit logarithmic increase with $t$ (through the term in $T$) that
is not very significant (it is in fact strictly zero when
$\delta/\alpha=1$). 
The true entropy is increasing in direct
proportion to the mass as is seen from the respective scalings. 

Let us
now substitute the coarse graining expansion from equation
(\ref{eq:1stcoarse}) to find to first order

\bea
{\cal S}&=&{\cal M}(\ln{4\pi^2}+(1-\delta/\alpha)\alpha T)+\nonumber\\
& &\int~dX~X^{2(1-\alpha/\delta)}~d\zeta~P_{oo}\left(\ln{1/P_{oo}}+(\alpha/\delta-1)\ln{X}\right)\nonumber\\
&+&\int~dX~X^{2(1-\alpha/\delta)}~d\zeta~\frac{P_{11}}{\alpha}X^{-\alpha/\delta}\left((\alpha/\delta-1)\ln{X}+\ln{1/P_{oo}}-1\right).
\label{eq:xent}
\eea

We consider first the undisturbed self-similar state wherein
$P_{11}=0$. The integrals over $\zeta$ and $X$ are independent (when
$\alpha/\delta>1$ the integral over $\zeta$ goes from
$-\sqrt{2\gamma}$ to $+\sqrt{2\gamma}$, when $\alpha/\delta<1$ it
varies from an arbitrary negative number to the same positive number)
so we may treat them as products. We observe that when
$\alpha/\delta>1$ the mixing part of the entropy (i.e. the integral
over phase space) has a well defined value provided that
$\alpha/\delta<3/2$. However this mixing entropy should be positive
and since the integral over $X$ is dominated by the behaviour at small
$X$ when $\alpha/\delta >1$ we see that this requires roughly that 
\be
X^{(\alpha/\delta-1)}>\exp{-\left(\int~d\zeta~P_{oo}\ln{1/P_{oo}})/\int~d\zeta~P_{oo}\right)}.
\label{eq:cutoff}
\ee
In other words, as is usual we expect that there will be an inner limit to the
extent of the steep self-similar state after which distortions must
arise. 

When $\alpha/\delta<1$ the negative mixing entropy will arise at large
$X$ so that the same inequality (\ref{eq:cutoff}) yields an upper
limit to the extent of the self-similar state. In general this limit
may overlap with the non-self-similar starting conditions for the
accretion of a dark matter halo. 

In the limiting case with $\alpha/\delta=1$ (which by equations
(\ref{eq:scaledens}) and (\ref{eq:denszero}) has the density profile
of the singular isothermal sphere), we see that the self-similar
entropy is strictly constant in the scaled variables. This suggests a
certain stability for this profile, but if we look at the
coarse-grained mixing entropy we see that it diverges for an infinite
system. Consequently such a state can never exist over all space for
this reason as well of course because of the infinite mass that would
imply. 
It does suggest that this might be the most stable self-similar state
in a finite relaxed region however \cite{LB67}. 

We turn now to the fine graining term involving $P_{11}$. This term
represents a departure from the strict self-similar density profile as
we have seen. Given however that it represents a finer graining of the
system, we should expect its contribution to the entropy to be
negative. Assuming that we are at small enough $X$ (the fiducial $X$
may be taken as the boundary of the core and the integral over $X$ in
this case is dominated by values at small $X$) that the bracket in
the first order term of equation (\ref{eq:xent}) is negative, then we
require $I_{11}>0$ for a negative first order term. By equation
(\ref{eq:P11}) if $P_{oo}$ is symmetric in $\zeta$ (such as is  a power
law in $\epsilon_{oo}$) then a sufficient condition is that the perturbed $P_{oo}$ satisfy 
$P_{oo}\propto |\epsilon_{oo}|^{(n)}$ where $n<1/2$. 
This last condition together with equations (\ref{eq:denszero}) and
(\ref{eq:potzero}) shows however that the disturbed coarse-grained
density will be flattened ($\theta_o\propto
X^{-2\alpha/\delta}~X^{(n-1/2)(1-\alpha/\delta)}$) when $n<1/2$. Thus
  {\it the deviation from strict self-similarity will be such as to flatten
  the steep self-similar density profile from the centre outwards}. One expects the
disturbances to arise near the centre of the system by our
considerations of the coarse-grained entropy.

For $\alpha/\delta<1$, the integral over $X$ is dominated by the
behaviour at large $X$. Thus the requirement for a negative
first order entropy (assuming the term in $\ln X$ to be dominant and negative)
is now $I_{11}<0$, and hence following the argument above and
referring to equation (\ref{eq:P11}) we see that once again
$n>1/2$. However an examination of the dependence of $\theta_o$ on $X$
as above now shows that the profile will be steepened. In this case we
can expect disturbances to arise at large $X$ by our coarse-grained
entropy argument. That is {\it the deviation from strict self-similarity for the
shallow initial profile is such as to steepen the profile from the
outside inwards}.
 This steepening is observed in the simulations to continue until the $r^{-2}$
 profile is obtained.  

Thus this exploration of the entropy function suggests that the flat
self-similar behaviour (as determined by initial conditions) is unstable to the development of steeper
behaviour , while the steep
self-similar behaviour is unstable to flattening in the central
regions. Ultimately incompleteness dominates both at the
centre and at the edge of the system as discussed above. Although this
predicts no real `universal' profile in the sense of \cite{SW98}, it does yield a universal qualitative behaviour in three
sections: wherein the middle section is close to a profile of
$r^{-2}$, an outer incomplete section is more like $r^{-3}$, and an
inner section that is flatter than $r^{-2}$ for a variety of reasons.  This may be in
accord with the simulations if not the observations. 

We should also observe that Evans and Collett \cite{EC97} have found
a remarkable self-similar steady state solution to the collisional
Boltzmann system that is an attractor and yields a flat $r^{-4/3}$
cusp. This is suggested 
to be applicable to galaxy formation by way of mergers where the
collisional ensemble is the set of merging clumps. Our approach
focusses on the collisionless dark matter particles and  stars and is therefore
independent of this result. However it would be interesting to apply 
the present method of coarse graining to a collisional system to see
how the Evans and Collett solution emerges. 

The analysis of this section has served primarily to test our coarse
graining expansion against relatively well known results, although the
results regarding the inevitable deviation from strict self-similarity
are new. In the next section we turn to the more challenging problem
of coarse graining the DF for a spherical system in self-similar
evolution  with velocity space anisotropy. That is $j^2\ne 0$ for each
particle, but there is no net rotation of the system \cite{HW95}.      

\section{Coarse Graining of Spherical Self-Similar Infall with
  Elliptical Orbits}
\label{sec:ellorb}
In this section we show how our procedure may be applied to more
complex systems. In principle we can dispense with spherical symmetry
and consider axially symmetric and three dimensional systems. However
the Green function solution of Poisson's equation then leads to algebraic
complexities that tend to obscure the method somewhat. This will be
attempted elsewhere. The present example is already new and physically
interesting.

The fundamental equations can be written in the form after Fujiwara \cite{Fuji83}
\bea
\partial_tf+v_r\partial_rf+\left(\frac{j^2}{r^3}-\partial_r\Phi\right)\partial_{v_r}f&=&0,\label{eq:DF}\\
\partial_r(r^2\partial_r\Phi)&=& 4\pi^2G\int~dj^2\int~fdv_r,\label{eq:pot}
\eea

where the particle density is given by 
\be
\rho=\frac{\pi}{r^2}\int~dj^2\int~fdv_r.
\label{eq:density}
\ee

We now define the anisotropic analogue of the self-similar radial
infall of the previous section. The definitions of $X$, $Y$ and $T$
are as given previously, while we introduce the extra scaled phase
space variable $Z$ according to 
\be
Z\equiv j^2e^{-\lambda T},
\label{eq:Z}
\ee
where on dimensional grounds we require 
\be
\lambda=4\delta-2\alpha.
\ee
In addition we use the scaled potential $\Psi$ as in equation
(\ref{eq:radpot}) while the scaled DF is written as 
\be
P(X,Y,Z) \equiv 4\pi^2 G fe^{(3\delta/\alpha-1)\alpha T}.
\label{eq:DFaniso}
\ee

Then the equations that define an anisotropic self-similar infall
model (ASSIM)are 
\be
-(3\frac{\delta}{\alpha}-1)P+(\frac{Y}{\alpha}-\frac{\delta}{\alpha} X)\partial_X
P-\left((\frac{\delta}{\alpha}-1)Y+\frac{1}{\alpha}(\frac{d\Psi}{dX}-\frac{Z}{X^3})\right)\partial_YP-(4\frac{\delta}{\alpha}-2)Z\partial_ZP=0,
\label{eq:SDFaniso}
\ee

and 
\be
\frac{d}{dX}\left(X^2\frac{d\Psi}{dX}\right)=\int~dZ~dY P.
\label{eq:SPoissaniso}
\ee
Moreover for the same scaling as in (\ref{eq:radrho}), the scaled
density becomes 
\be
\theta =\frac{\pi}{X^2}\int~dZ~dY P.
\ee
Once again the coarse graining consists in expanding all quantities as
in equation (\ref{eq:expansion}).

In this case we have scaled the phase space volume to 
\be
\Delta r\Delta v_r\Delta j^2\equiv \Delta X\Delta Y\Delta Z
e^{(6\delta/\alpha-3)\alpha T}
\label{eq:anphase}
\ee
so that the condition for coarse graining at fixed $T$ in the limit
$\alpha\rightarrow\infty$ remains $\delta/\alpha>1/2$. One should note
that should $\delta/\alpha< 1/2$, the expansion is about the
fine grained limit since infinite $\alpha$ yields $P_o$ as the exact
result.

Proceeding with the expansion the zeroth order equations become;
\bea
(3\frac{\delta}{\alpha}-1)P_o+\frac{\delta}{\alpha}X\partial_XP_o+(\frac{\delta}{\alpha}-1)Y\partial_YP_o+(4\frac{\delta}{\alpha}-2)Z\partial_ZP_o&=&0\\
\frac{d}{dX}\left(X^2\frac{d\Psi_o}{dX}\right)&=&\int~dZ~dY P_o.
\label{eq:anzeroP}
\eea
We can solve for $P_o$ by the method of characteristics to find 
\be
P_o=P_{00}(\zeta_1,\zeta_2)e^{-(3\delta/\alpha-1)s},
\label{eq:Pzero}
\ee
where the characteristic constants are 
\bea
\zeta_1&\equiv& Y/X^{(1-\alpha/\delta)}\label{eq:onezeta}\\
\zeta_2^2&\equiv& Z/X^{(4-2\alpha/\delta)}\label{eq:twozeta},
\eea
and the arc-length along the characteristic $s$ may be taken as
in (\ref{eq:arc}).

It is interesting already to note that the scaled density may be
written as 
\bea
\theta&=&\pi X^{-2\alpha/\delta}I_{oo}\\
      &\equiv&\pi X^{-2\alpha/\delta}\int~P_{oo}(\zeta_1,\zeta_2)d\zeta_1d\zeta_2^2,
\label{eq:anisodens}
\eea
which shows that (provided the integral $I_{oo}$ goes over the same set of
characteristics at every $X$) the usual self-similar density profile
holds even to zeroth order, just as for the radial orbits. Since there is no potential term in the
governing equations at this order, this must be set solely by the
initial conditions, which determine the similarity class
$\alpha/\delta$. Consequently, the zeroth order potential in this case
is as in equation (\ref{eq:potzero}) except that the integral $I_{oo}$
is here defined as the integral over $\zeta_1$ and $\zeta_2$ that
appears above in the scaled density.

Using the same notation as in the radial case, the scaled energy is 
\be 
\epsilon =\Psi+\frac{Y^2}{2}+\frac{Z}{2X^2},
\ee
and so in zeroth order with
$\epsilon_o=\epsilon_{oo}X^{(2-2\alpha/\delta)}$ as for radial orbits  
\be 
\epsilon_{oo}=-\gamma +\frac{\zeta_1^2}{2}+\frac{\zeta_2^2}{2},
\ee
where $\gamma$ is defined as before in terms of the current $I_{oo}$.  

Proceeding to the first order in the expansion we find the governing
equations to be
\bea
\frac{\delta}{\alpha}X\partial_XP_1+(\frac{\delta}{\alpha}-1)Y\partial_YP_1+2(\frac{2\delta}{\alpha}-1)Z\partial_ZP_1&=&-(\frac{3\delta}{\alpha}-1)P_1\nonumber\\
&+&Y\partial_XP_o-\left(\frac{d\Psi_o}{dX}-\frac{Z}{X^3}\right)\partial_YP_o,\\
\frac{d}{dX}\left(X^2\frac{d\Psi_1}{dX}\right)&=&\int~dZ~dY P_1.
\eea  
Thus the form of the characteristics remain the same as in equations
(\ref{eq:onezeta}), (\ref{eq:twozeta}) and the solution for $P_1$ by
the method of characteristics yields (plus a term that may be absorbed
into the zeroth order since it has the same dependence on $s$)
\bea
P_1&=&-e^{-3\frac{\delta}{\alpha}s}P_{11}(\zeta_1,\zeta_2)\nonumber\\
   &=&-X^{-3}P_{11}(\zeta_1,\zeta_2),
\label{eq:First}
\eea
where 
\be
P_{11}\equiv
\left ((\frac{\alpha}{\delta}-1)\zeta_1^2-\frac{I_{oo}}{(3-2\alpha/\delta)}+\zeta_2^2\right)\partial_{\zeta_1}P_{oo}+2(\frac{\alpha}{\delta}-2)\zeta_1\zeta_2^2\partial_{\zeta_2^2}P_{oo}-(3-\frac{\alpha}{\delta})P_{oo}.
\ee

Consequently the solution to first order in the coarse graining
parameter for the DF is 
\be
P=P_{oo}(\zeta_1,\zeta_2)X^{-(3-\alpha/\delta)}-\frac{1}{\alpha}X^{-3}P_{11}(\zeta_1,\zeta_2),
\label{eq:anisoDF}
\ee
and the corresponding density profile becomes formally that of
equation (\ref{eq:1stcoarse}) but for a numerical factor, namely  
\be
\theta=\pi
X^{-2\alpha/\delta}\left(I_{oo}-\frac{1}{\alpha}I_{11}X^{-\alpha/\delta}\right),
\label{eq:andensity}
\ee
Here however the integrals are defined as
$I_{ii}=\int~P_{ii}d\zeta_1d\zeta_2^2$.

We arrive then at a conclusion similar to the one found for purely
radial orbits, namely that deviations from the self-similar density
profile must arise as $X\rightarrow 0$. With $P_{11}>0$ for example so that
$I_{11}>0$, we see that the profile will be flattened near the centre
of the system as the resolution increases. Should $P_{11}=0$ then we
have the condition on $P_{oo}$ that the system remain self-similar at all $X$ to
all orders. This requires redefining $P_{oo}$ to be
$P_{oo}+\alpha^{-1}P_{o1}+\dots$ where the $P_{oi}$ are the
contributions to the zeroth order variation from the $i^{th}$
order. If these functions are taken to be zero, the argument is exact.

The argument based on entropy (that we do not reproduce
here for reasons of brevity) suggests that there should be flattening
at the centre and so we turn to consider the condition $P_{11}\ge
0$. Just as in the radial case we may hope to establish the DF that
maintains the coarse-grained self-similarity exactly and to put some
constraints on deviations that maintain $P_{11}>0$.

The equation $P_{11}=0$ is a partial differential equation for the
exact self-similar DF. We may solve this equation once again by the
method of characteristics to find 
\be
P_{oo}=F(\kappa)e^{(3-\alpha/\delta)\ell},
\label{eq:SSanDF}
\ee
where 
\be
\kappa\equiv
\left(\frac{\zeta_1^2}{2}+\frac{\zeta_2^2}{2}-\gamma\right)(\zeta_2^2)^{-b},
\label{eq:kappa}
\ee
and setting $v\equiv \zeta_2^2$ for convenience 
 \be
\pm 2(\alpha/\delta-2)\frac{d\ell}{dv}=\frac{1}{v\sqrt{2\kappa
    v^{2b}-v+2\gamma}}.
\label{eq:ell}
\ee

In these formulae the constants are 
\bea
b&\equiv& \frac{\alpha/\delta-1}{2(\alpha/\delta-2)}\nonumber,\\
\gamma&\equiv& \frac{I_{oo}}{2(3-2\alpha/\delta)(\alpha/\delta-1)}.
\eea

The preceding solution holds for $\alpha/\delta\ne 1$ and in that
special (isothermal) case we find 
\be
P_{oo}=F(\kappa)e^{2\ell},
\ee
where now 
\be
\kappa\equiv
\frac{\zeta_1^2}{2}+\frac{\zeta_2^2}{2}+\frac{I_{oo}}{2}\ln{\zeta_2^2},
\ee
and with $v$ as above
\be 
\frac{d\ell}{dv}=\pm \frac{1}{2v\sqrt{2\kappa-v-I_{oo}\ln{v}}}.
\ee
Now it is readily found by tracing back through the various
definitions that this coarse-grained self-similar or power law
solution is a steady state. And in fact we have the relations 
\bea
\zeta_1^2&=&(v_r^2)r^{-2(1-\alpha/\delta)},~~~
\zeta_2^2=j^2r^{-(4-2\alpha/\delta)},\nonumber\\
\kappa&=&\left(\frac{v_r^2}{2}+\frac{j^2}{2r^2}-\gamma\right)(j^2)^{-b},~~~f=\frac{1}{4\pi^2G}r^{-(3-\alpha/\delta)}F(\kappa)e^{(3-\alpha/\delta)\ell}.\nonumber 
\eea

We can observe as a kind of verification of our procedure that
in the special case $\alpha/\delta=3$, the distribution function is
simply $f=f(\kappa)$. In fact it is a member of the steady state solutions found
for the anisotropic spherically symmetric case in \cite{HW95}, and
used previously to model galaxies by Kulessa and Lynden-Bell
\cite{KL92}, although it is one of the unbound cases. It appears here
in zeroth order because in this case as may be seen from
equation (\ref{eq:anphase}) our expansion is actually a {\it fine
  graining} expansion for $\alpha/\delta>2$. This means that the
zeroth order is the exact solution as $\alpha\rightarrow\infty$. For
$\alpha/\delta<2$ we appear to have another set of steady solutions
which develop as limits from the time dependent system.  

The function $\ell(v)$ is not readily found analytically but it is in
general an elliptic function. This function serves to establish a
finite range in permitted values of $v\equiv \zeta_2^2\equiv
j^2/r^{-(4-2\alpha/\delta)}$. Thus for example when $\alpha/\delta=1$
we see that all large $r$ at a given $j^2$ is permitted, but small
enough $r$ is forbidden. This is as expected. Similarly if
$3/2>\alpha/\delta>1$ so that $\gamma$ and $b$ are positive but
$\kappa$ may be negative, or if $\alpha/\delta<1$ so that $\gamma$ is
negative but $b$ and $\kappa$ are positive; then small $r$ is also
excluded for a given $j^2$.   
 
An obvious application of this coarse graining procedure is to the
virialized core that
develops from a recollapsing dark matter halo. We do not now 
impose that the high order terms in the coarse graining
expansion vanish. In the self-similar infall model of dark matter halo
formation 
(e.g. \cite{HW99}) the similarity `class' $\alpha/\delta$ is given by 
\be
\alpha/\delta=\frac{3\epsilon}{2(\epsilon+1)},
\label{eq:sclass}
\ee
where $-\epsilon$ is the index of the density power law in the
primordial density fluctuation. From equation (\ref{eq:andensity}) we
find 
\be
\frac{d\ln\theta}{d\ln{X}}=-\frac{2\alpha}{\delta}+\frac{\alpha}{\delta}\frac{(I_{11})X^{-(\alpha/\delta)}/\alpha}{I_{oo}-(I_{11}X^{-(\alpha/\delta)}/\alpha)}.
\ee
Thus for example with $\epsilon=2$ or $\alpha/\delta=1$ and allowing
the first order term $I_{11}X^{-(\alpha/\delta)}/\alpha$ to be as
large as $(1/2)I_{oo}$, we see that the normal index of self-similar infall
has flattened to $-2+1=-1$. Thus the flattening at or inside this
radius (where higher order terms in the coarse graining expansion must be taken
into account)is in agreement with the NFW profile. However such a
value for $\epsilon$ is only reasonable on the largest halo scales. On
the scale of galaxies the observed power spectrum index of $n\approx
-2$ can be interpreted (\cite{HW99}) as requiring $\epsilon =1/2$ and
hence $\alpha/\delta=1/2$. In this case the central flattening changes
the self-similar index of $-1$ to flatter than $-1/2$ inside the same
radius as above.

 \section{Numerical Experiments}
\label{sec: numexp}

In what follows we use a second generation shell code that is written
in the scaled variables that were used in the previous sections \cite{LeD01},
\cite{HW99} to test the predictions of the section on radial orbits.
The details of the code may be found in \cite{LeD01}, but it is
important to realize that these results are based in part (the DF figures) on an analytical
estimate of the core particle potential energy in an effort to suppress
the noise in a small number shell code simulation. This allows us to
take the core shell potential energy to be proportional to $-GM(r)/r$ since
the additional term in the potential energy $4\pi
G\int_{r_c}^r~\rho(r')r'dr'$ (where $r_c>0$ is a central reference
radius) is itself proportional to $GM/r$ when the analytic expression
for the density is used. Since the eventual steady state of the core
appears to be reflected first in the density profiles, this seems to
be a reasonable way to remove the noise that is due to shell crossing and arises during a discrete
evaluation of the integral term . We are in fact able to reproduce known steady
state distribution functions by using this technique. The second noise
reducing ingredient is the addition of a central point mass, which
assists the density profile to obtain its stable values. As a result
of the rapid establshment of a steady core we are also able to reduce
noise in the PDF by averaging over different stable epochs.   
 
 We turn first to the question of whether in the radial self-similar
 infall phase the DF of a dark matter halo can really be taken as the
 `one-half law`
 (\ref{eq:canonDF}). To this end we have simulated the standard
 cosmological model of radial self-similar infall (e.g.\cite{FG84}; \cite{B85}; \cite{HW99}; and references
 therein) of collisionless matter  using  10,000 equal mass spherical shells. We have found that the
 self-similar `equilibrium' phase is greatly stabilized
 microscopically by the addition of a central point mass. This point
 mass is negligibly small ($\approx 5\times 10^{-4}$) compared to the mass of the halo so that it
 does not affect the global dynamics, but it imitates a central
 density cusp of zero mass as found analytically in the self-similar
 phase. This can not be realized numerically otherwise. It is
 particularly effective in the `shallow' case where the halo mass is
 going to zero with the radius.

We consider the two cases referred to as `flat' or `steep' in the
relevant literature depending on whether the initial cosmological
density profile is flatter or steeper than $r^{-2}$. The steep cases
all achieve an intermediate self-similar phase wherein the stable density profile is given by
the initial conditions ; while the flat cases, although growing
self-similarly, all establish a `universal' $r^{-2}$ density
profile in the intermediate self-similar phase. That is, the $r^{-2}$ profile serves as a one-sided
`attractor' for the self-similar infall. The cases that we show here
however are selected for their stability rather than to show the
evolution towards the attractor. This evolution has been demonstrated
elsewhere at length (\cite{HW99} and references therein). We are
primarily interested here in the stable density and PDF profiles. 

\begin{figure}
{\par \centering\epsfig{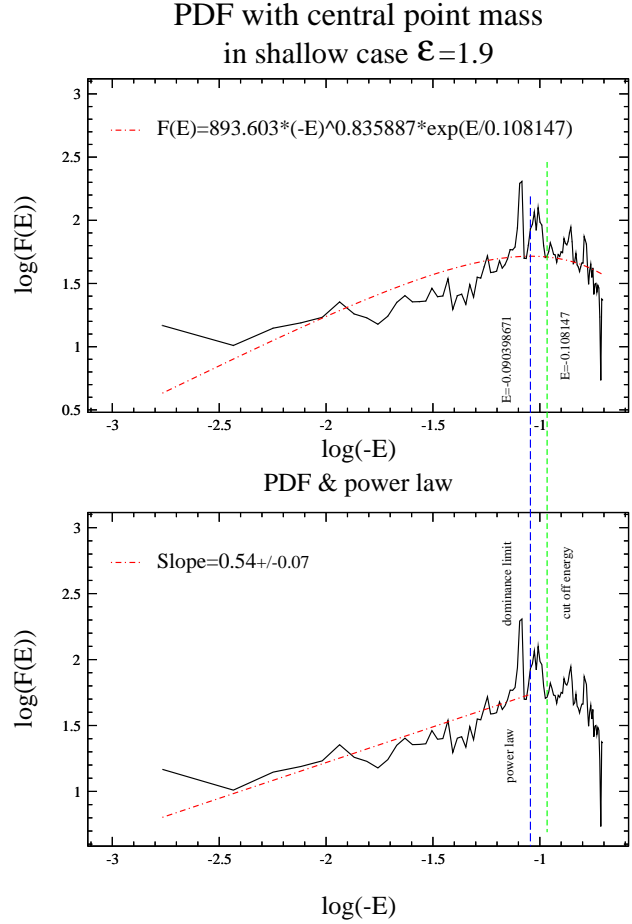}\par}
\caption{In both panels the data give the PDF of the simulation for a
  shallow initial density profile ($\epsilon=1.9$) when averaged over
  the self-similar phase. The simulation was run with a small central
  point mass ($~5\times 10^{-4}$ the total mass of the halo) that substantially reduces the noise in the core without
  affecting the global behaviour.  The upper panel shows an attempt to
  fit a power law times an exponential, while the lower panel fits a
  power law up to the point where the curvature may not be ignored. }
\label{fig:flatPDF}
\end{figure}

In figure (\ref{fig:flatPDF}) we show the PDF averaged over the
self-similar phase of the simulation for a `slightly flat' case where
the initial power law is $\propto r^{-1.9}$. The expected core profile
in this case is that of the attractor, namely $r^{-2}$.In the top panel we
attempt to fit a negative temperature gaussian multiplied by an energy
power law. This serves mainly as an indicator (at the maximum of the
curve) of the limit to which the power law fit of the lower panel may
be extended. This is the `dominance limit' indicated on the figures. Such a law was suggested in \cite{MTJ89} and subsequently in \cite{HW99} for
radial infall with the power equal to $1/2$. We see from the figure
that this does not provide a very good fit to the simulated PDF. But
if we focus only on the region where a simple power law may apply as
in the lower panel, then we find good evidence for the expected
behaviour 
(the slope of the averaged PDF over nearly one and one half
decades is $0.54\pm .07$). The eventual steep cut-off at large negative
energies may arise in part  due to the finite nature of the simulation
wherein the most negative energies are cut off at large radii, but the
preceding curvature may be  evidence of a central (assuming that the
most negative energies are now near the geometrical centre) gaussian PDF. We
would expect such a cusp based on the arguments of section
(\ref{sec:Coarse Grain}), to the extent that the averaged behaviour
approximates a steady state.  



\begin{figure}
{\par \centering\epsfig{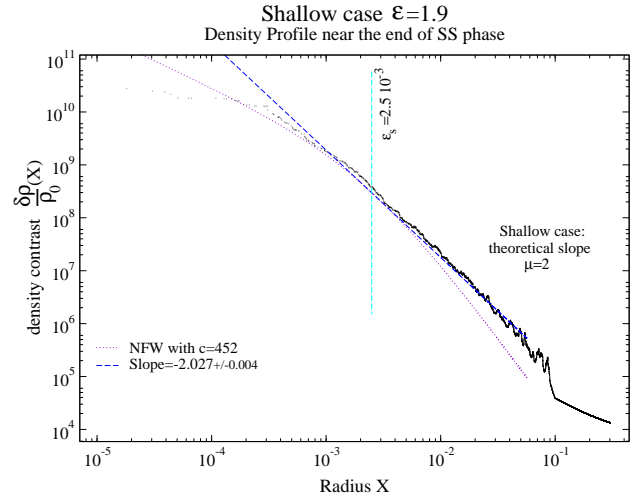}\par}
\caption{Density Profile established near the end of the Self-Similar
  phase for an initially slightly flat perturbation. The smoothing
  scale is shown by a vertical line, but the straight line fit with
  slope $-2$ extends to the interior of this scale. The particles in
  the outer mixing zone of the halo have been neglected. The NFW
  profile provides a poor fit in this purely radial case. }
\label{fig:flatdens}
\end{figure}

In figure (\ref{fig:flatdens}) the density profile established in the
dark matter core near the end of the self-similar phase is shown. We
see principally that the theoretical attractor profile of $r^{-2}$ is quite
convincingly established over an intermediate range of scales that
extends to the interior of the smoothing scale. We take this to be
confirmation of the idea that the main relaxation of the core occurs
near the boundary turning point since in this case more than the
initial conditions are required to achieve the attractor profile as
found.

The profile steepens at the outside due partly to the absence of particles
( the `Keplerian effect' see e.g. \cite{HW99}) and
partly perhaps due to the fine grained entropy effect (the outer parts
of the core retain the pure phase mixing ) suggested at the end of the
 section on radial orbits. The flattening at small radii here is largely due to  the dominance of the point mass
potential. For this purely radial model with a central point mass the NFW profile
provides a poor fit as shown.     
    
\begin{figure}
{\par \centering\epsfig{file= 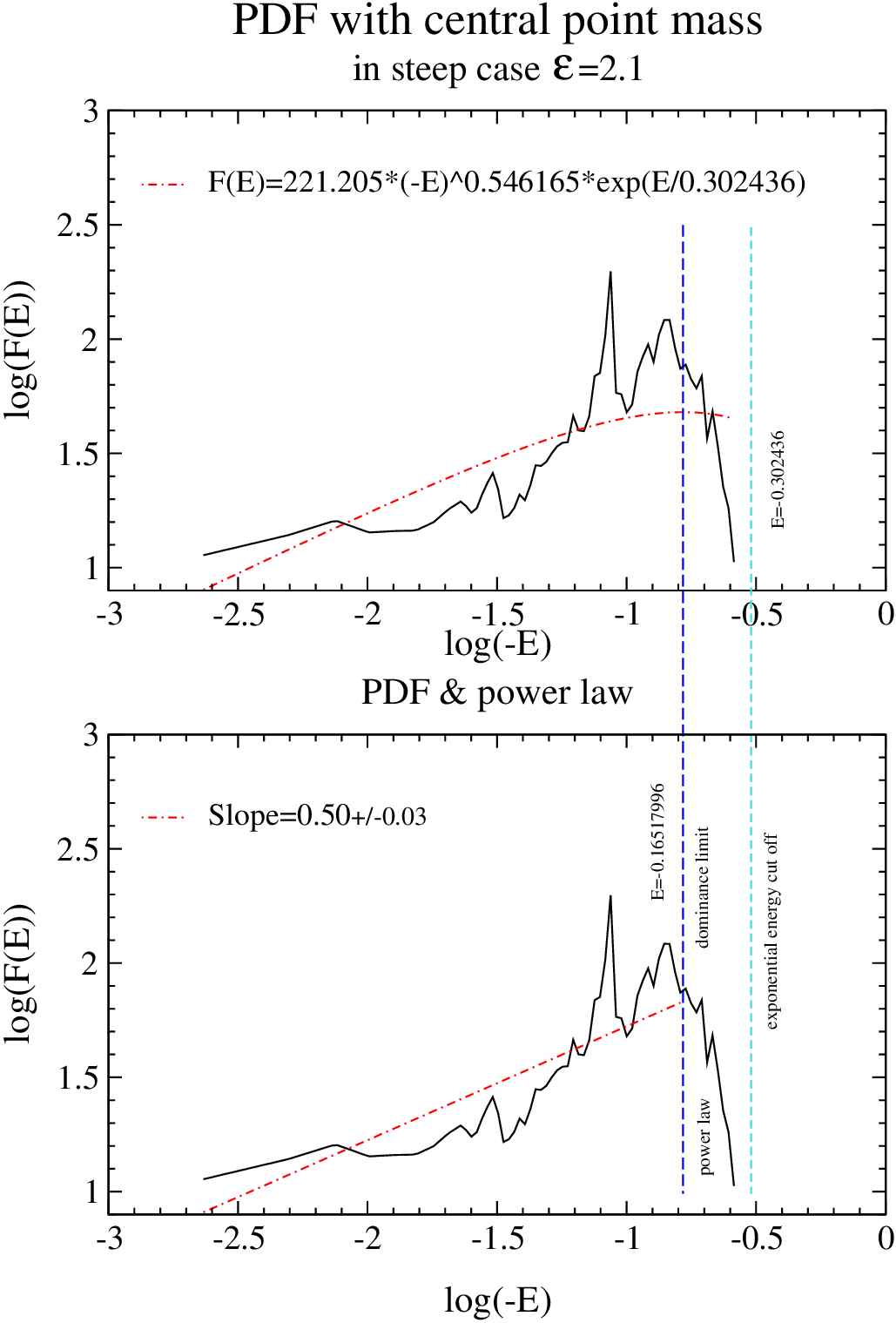,width=0.5\linewidth}\par}
\caption{In both panels the data give the PDF of the simulation for a
  steep initial density profile ($\epsilon=2.1$) when averaged over
  the self-similar phase. The simulation was run with a small central
  point mass ($~5\times 10^{-4}$ the total mass of the halo) that substantially reduces the noise in the core without
  affecting the global behaviour.  The upper panel shows an attempt to
  fit a power law times an exponential, while the lower panel fits a
  power law up to the point where the curvature may not be ignored. }
\label{fig:steepPDF}
\end{figure}

\begin{figure}
{\par \centering\epsfig{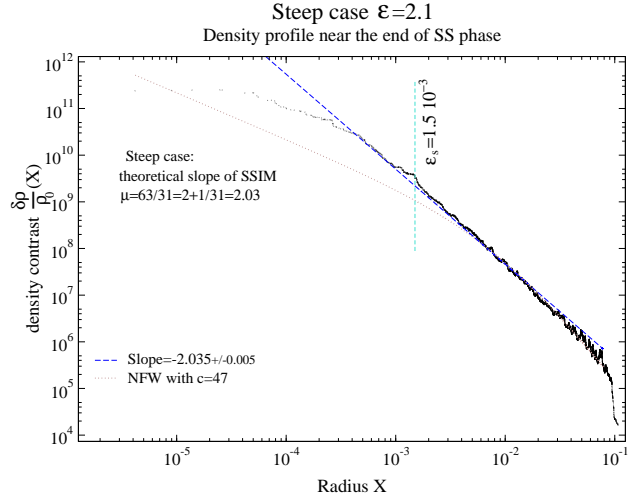}\par}
\caption{Density Profile established near the end of the Self-Similar
  phase for an initially slightly steep perturbation. The smoothing
  scale is shown by a vertical line, but the straight line fit with
  slope $-2.03$ extends to the interior of this scale. The particles in
  the outer mixing zone of the halo have been neglected. The NFW
  profile can provide a good fit to the outer part of the halo but it
  is far too flat in the inner regions for these purely radial orbits. }
\label{fig:steepdens}
\end{figure}

In figures (\ref{fig:steepPDF}) and (\ref{fig:steepdens}) the same information as above for a
`slightly steep' case is presented (the expected slope of
$2\alpha/\delta =2.03$ by equation (\ref{eq:sclass}).

 The PDF is again shown fitted
by a negative temperature exponential times a power and by a pure
power law. The power law fit yields $0.50\pm .03$ when it is taken
from the outer core regions to the maximum of the exponential
curve, some one and one-half decades. However it is a poor detailed
fit to the data here and must be considered a weaker result than is
the shallow case. The steep cut off at high negative energies probably comes in
this case  from the
finite numerical resolution of the central regions (where now the most
negative energy particles initially, originate), but once again there is a hint
of a gaussian just before this cut-off.

Figure (\ref{fig:steepdens}) shows that the steep simulation has
reproduced the expected density profile in the intermediate scales
rather accurately, and that the profile continues well inside the
smoothing length as anticipated. There is also the outer Keplerian
steepening  and an inner flattening due to the potential of the
central point mass. Once again no globally good fit with the NFW
profile is possible.  

We believe that these numerical experiments offer some confirmation of
the ideas expressed earlier in this paper and in \cite{HW99}. In particular the radial
PDF is more strongly suggested to be $(-E)^{1/2}$ in the relaxed self-similar
region. This cuts off at large negative energies due to a combination
of initial conditions and finite mass resolution, but before this
occurs, there is slight evidence for a Gaussian form in the central
regions as suggested by our coarse graining in section (\ref{sec:Coarse Grain}). 

The density profiles agree with previous work in the
self-similar infall phase, but we have reduced the noise in the
simulation by the device of including a central point mass of
negligible size. The fact that the slopes continue substantially
inside the smoothing length suggests that the key behaviour
determining the profile occurs in the mixing region near the boundary
of the core (and $v=0$) combined with the initial conditions. This
also agrees with and supports our coarse graining argument of section
(\ref{sec:SScoarse grain}). It is also clear from these figures that
the NFW profile can not provide a global fit to these radial infall
models. This then requires an exploration of the effects of angular
momentum, which we reserve for the next paper in this series. 
 
\section {Conclusions and Summary}

Our main interest in this paper has been to test the usefulness of a
proposed coarse-graining (actually a progressively finer grained series) technique that uses non-canonical coordinate
transformations on the phase space of a dynamical system. This was
used in two cases in sections (\ref{sec:Coarse Grain}) and
(\ref{sec:SScoarse grain}) respectively. The first section concerned
itself with the coarse graining of an equilibrium, spherically
symmetric system by a stretching transformation on phase space. This
allowed us to conclude that near the centre of such a system 
regularity of the coarse graining expansion imposed some conditions on
the coarse-grained $f(v^2)$. The one which maintained its form to all
orders and which gave the same form at all sufficiently small $r$, was
a Gaussian. We also checked that the entropy decreased in the higher
orders (finer grained) of the expansion without imposing further
constraints.
The coarse graining expansion led naturally to the small $r$ expansion
for the density profile that is familiar from the theory of polytropes.

In the second section we treated the radial self-similar infall model
that is of significance for cosmological dark matter halos. The
coordinate transformation used for the coarse graining was also the
transformation that renders the system stationary. We found that the
coarsest grained  density profile was already that dictated by the
self-similarity class implicit in the initial conditions. Some
evidence for this behaviour was found in the numerical simulations of  
section (\ref{sec: numexp}) in that the density profile extended well
inside the smoothing length of the simulation. We concluded that the
actual relaxation of the system took place near the `boundary' of the
core.  

The requirement that the self-similarity be stable in these
coordinates to all orders of the coarse-graining expansion (ie
essentially that the coarse grained and fine grained DF's be equal) yielded the
self-similar DF as $f\propto \delta (j^2)(|E|)^{1/2}$. This had been
previously conjectured in \cite{HW99}, but the present argument is
more direct. Moreover from our simulations the evidence presented in
support of this result appears stronger than before, although we
incorporate a partly analytical estimate of the shell energy here
rather than use the much noisier numerical calculation. Had we not
done this our evidence would have been weaker than in \cite{HW99}. 
We also found
that the presence of a point mass at the centre of the system greatly
reduced the noise throughout the core in our simulations.

Eventually the DF is cut off in the central regions of the system
(large negative energies) due to finite mass resolution and/or initial
conditions. There is weak evidence for a Gaussian in velocity (an
exponential in  energy) before this cut-off, as might be expected from
the work of section (\ref{sec:Coarse Grain}). This is because the
central regions of the core resemble a steady state as discussed in
\cite{HW99}, due probably to the relaxation time being short compared
to the `accretion time' (time to significantly change the mass) in
this region. 

By expressing the entropy in the scaled variables and requiring the
coarse-grained entropy to be positive, we saw that there was an inner
limit to the self-similarity for the cosmologically `steep'
case. Similarly there was an outer limit to the extent of the
cosmologically `flat' case. The limiting case which has an inverse
square density profile has an exactly constant entropy in self-similar
variables that is always positive. It diverges for an infinite system
but it is likely to be the most stable intermediate profile for finite
systems.

By requiring the next finer grained entropy contribution to the
total entropy to be
negative (corresponding to the increased information at this order) we
inferred that the initially steep self-similar infall should flatten,
primarily at the centre of the system, while the initially flat
self-similar infall should steepen, primarily near the exterior. In
this way an intermediate attractor behaviour exists in the
self-similar infall model. However the behaviour at small radii is
sensitive to the presence of point masses while the behaviour at large
radii (exterior to most of the mass) tends to be Keplerian
(\cite{HW99}). Thus the global profile is more subtle in the manner of
the NFW fit. However these purely radial systems are NOT well fitted
by the NFW profile. 

We also studied theoretically (but not numerically) the spherically symmetric infall model in the presence
of non-zero angular momentum. We showed that the coarse graining
expansion can be carried out in just the same way as for radial
orbits. The expected density profiles are not changed and flattening
is expected in the central regions. The flattening is of a progressive
nature and becomes even flatter than $r^{-1}$ of the NFW profile as $r\rightarrow
0$. This is compatible with the expected Gaussian DF in the central
regions of the system, although this can only be established
definitely by examining higher orders in the expansion. There is an
interesting dependence of the expected outer power law and flattening
on scale through the index of the initial density perturbation
(related to the power spectrum index $n$). This is such as to predict
flatter cusps for galaxies than the NFW profile, which in turn is most
appropriate on cluster scales or above ($n=1$).

We were able also to constrain the form of the DF in the relaxed state
when the coarse grained and fine grained functions are the same as in
(\ref{eq:SSanDF}). These relaxed DF's are steady states as is the
radial equivalent. They appear to be consistent with a possible
Gaussian DF modulated by a function that allows for inner and outer
turning points at a given $j^2$. This is presumably due to the
somewhat artificial constraint of spherical symmetry, which restricts
the relaxation in angular momentum. 

We have found that the NFW profile can be produced in
such a spherical system with an appropriate distribution of angular
momentum (and no central black holes). However the angular momentum
has to be correlated with initial radius of a particle in a power law fashion that is
consistent with the self-similarity. Once again this may work only in
the absence of strong relaxation in angular momentum. We intend to
explore this question by coarse graining axially symmetric systems in
a subsequent paper.  

\section{Acknowledgements}
RNH acknowledges the support of an operating grant from the canadian
Natural Sciences and Research Council. M LeD wishes to acknowledge the
financial support of Queen's University.

\label{lastpage}    
\end{document}